\documentclass[twocolumn,showpacs,preprintnumbers,amsmath,amssymb]{revtex4}

\usepackage{graphicx}
\usepackage{dcolumn}
\usepackage{bm}

\begin{document}
\title{Roughness of stylolites: 
a stress-induced instability with non local interactions}

\author{J. Schmittbuhl} 

\affiliation{Laboratoire de G{\'e}ologie, UMR CNRS 8538, Ecole Normale
Sup{\'e}rieure,\\ 24, rue Lhomond, F--75231 Paris C{\'e}dex 05,
France. Email: Jean.Schmittbuhl@ens.fr.}

\author{F. Renard and J.P. Gratier} 

\affiliation{LGIT-CNRS-Observatoire, Universit\'{e} J. Fourier BP 53,
38041 Grenoble, France.}

\date{\today}
\begin{abstract} 
We study the roughness of stylolite surfaces (i.e. natural
pressure-dissolution surfaces in sedimentary rocks) from profiler
measurements at laboratory scales. The roughness is shown to be nicely
described by a self-affine scaling invariance. At large scales, the
roughness exponent is $\zeta_1 \approx 0.5$ and very different from
that at small scales where $\zeta_2 \approx 1.1$. A cross-over length
scale at around $\lambda_c =1$mm is well characterized and interpreted
as a possible fossil stress measurement if related to the
Asaro-Tiller-Grinfeld stress-induced instability. Measurements are
consistent with a Langevin equation that describes the growth of
stylolite surfaces in a quenched disordered material with long range
elastic correlations.
\end{abstract} 
\maketitle
Stylolites are geological patterns that are very common in polished
marbles, a material largely used for floors and walls of buildings and
monuments. They exist in many sedimentary rocks such as limestones,
sandstones or evaporites \cite{Bathurst71}. They are rock-rock
interfaces, that are formed during diagenesis and result from combined
stress-induced dissolution and precipitation processes
\cite{Dewers90,Ortoleva94,Park68}. They exist on a very large range of
scales, from micro-meters to tens of meters. They are often observed
as thin irregular interfaces that look like printed lines on rock cuts
which explain their name.  At larger scales, they are planar
structures that are typically perpendicular to the tectonic load
(i.e. lithostatic pressure). They might form complex network with or
without connections.

However, despite their abundance, stylolites are as mentioned by {\it
Gal et al.} \cite{Gal98} {\it ``among the least well-explained of all
pressure-solution phenomena''}. First they are complex 3D structures
that are often only described from 2D cross-sections. Second, they
develop in various petrological and tectonic contexts which lead to
very different geometries. Third they are often transformed and
deformed because of post-processes like diagenesis and metamorphism
that develop after their initiation.

In this letter we propose a new mechanism for stylolites growth that
is consistent with recent 3D roughness measurements of a stylolite
interface. The first part of the letter deals with the topography
measurement and its description in terms of scaling invariance, namely
self-affinity. The second part is devoted to a physical model of the
stylolites roughening. It is based on a Langevin equation that
accounts for a stress-induced instability in a quenched disorder with
capillary effects and long range elastic interactions.

\begin{figure}[h]
\begin{center}
\vspace*{5mm}\includegraphics[width=8cm,angle=0]{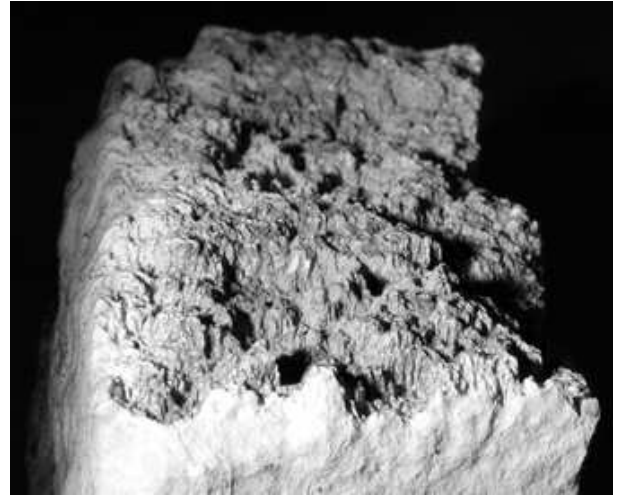}
\caption{\label{fig_stylo_sample}Picture of the stylolite surface that
has developed in a limestone from Juras Mountains (France). Magnitude
of the peaks are typically of the order of 2 mm. The width of the
sample is 4 cm.}
\end{center}
\end{figure}
The roughness measurement has been performed on a stylolite interface
included in a limestone sample from Juras Mountains in France (see
Fig.~\ref{fig_stylo_sample}).  The sample has been collected in a
newly open quarry, thus preserved from late breakage and chemical
erosion. The opening procedure was possible for this sample because of
the accumulation of undissolved minerals like clays that form a weak
layer along the stylolite interface. The concentration of these
minerals provides an estimate of the cumulative strain that the sample
has undergone \cite{Renard03}.  As shown in
Fig.~\ref{fig_stylo_sample}, peaks along the interface are randomly
distributed in space and of various sizes (up to several
millimeters). Locally slopes along the topography might be very
important which makes the roughness measurement difficult.

\begin{figure}[h]
\begin{center}
\includegraphics[width=8cm,angle=0]{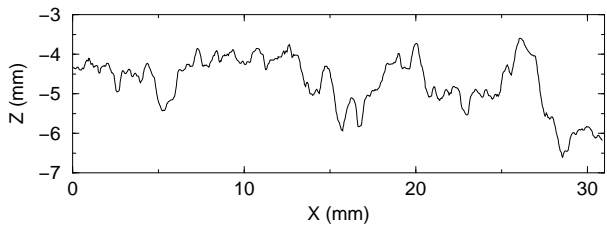}
\caption{\label{fig_roughness_prof} A 1D profile obtained by a
mechanical profiler (1030 data points -
$\Delta x=30\mu$m). }
\end{center}
\end{figure}
We used two different profilers to sample the stylolite
roughness. First, with a mechanical profiler
\cite{Schmittbuhl95,Lopez98} we extracted four profiles of 1030 points
each with a horizontal step of $\Delta x=30\ \mu$m. The specificity of
the mechanical profiler is to measure the surface height from the
contact of a needle onto the surface. The radius of the needle tip is
25 $\mu$m. The vertical resolution is $3\ \mu$m over the available
range of 5 $cm$. One of this profile is shown in
Fig.~\ref{fig_roughness_prof}. We compare the mechanical measurement
to an optical profiling \cite{Meheust02}. This technique is based on a
laser triangulation of the surface without any contact with the
surface. The laser beam is $30\ \mu$m wide. Horizontal steps between
measurement points were $\Delta x=\Delta y = 50\ \mu m$. The main
advantage of this technique is the acquisition speed that can be
significantly larger compared to the mechanical profiler since there is
no vertical move. However, a successful comparison with mechanical
measurement is necessary to ensure that optical fluctuations are
height fluctuations and not material property fluctuations.

\begin{figure}[h]
\begin{center}
\includegraphics[width=7cm,angle=0]{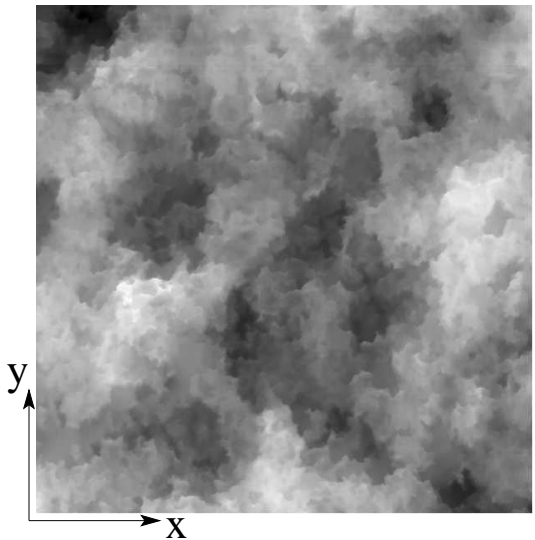}
\caption{\label{fig_roughness_surf} 2D map obtained by an optical
profiler (600$\times$600 data points - $\Delta x=\Delta
y=50\mu$m).  Heights are gray encoded:
white corresponds to the minimum height and black to the maximum
height.}
\end{center}
\end{figure}

Fig.~\ref{fig_roughness_surf} shows a gray level map of the stylolite
surface heights as a high resolution image ($600\times 600$ pixels)
that covers over a surface area of 3$\times$3 cm$^2$. The vertical
resolution is $2\ \mu$m over the $6$ mm vertical range of the setup.

\begin{figure}
\begin{center}
\includegraphics[width=8cm,angle=0]{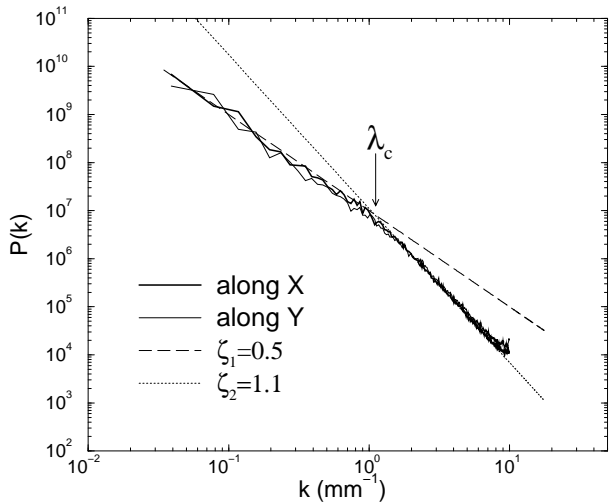}
\caption{\label{fig_spum_data} Averaged power spectrum of the
topographic profiles extracted from the optical map of the stylolite
surface.}
\end{center}
\end{figure}

We analyzed the height distribution in terms of self-affinity
\cite{Barabasi95} which states that the surface remains statistically
unchanged for the transform: $\Delta x \rightarrow \lambda \; \Delta
x$, $\Delta y \rightarrow \lambda \; \Delta y$, $\Delta z \rightarrow
\lambda^\zeta \; \Delta z$, where $\lambda$ can take any real
value. The exponent $\zeta$ is the so-called roughness exponent. A 1D
Fourier spectrum of a self-affine profile is shown \cite{Barabasi95}
to behave as a power law with a slope $-1 - 2 \: \zeta$, and provides
an estimate of the roughness exponent $\zeta$. We applied this
technique whose reliability has been tested \cite{Schmittbuhl95b}, to
profiles extracted from the optical map shown in
Fig.~\ref{fig_roughness_surf}, either along $x$ or $y$
directions. Results are plotted in Fig.~\ref{fig_spum_data}. The
figure shows with a thick solid line the $y$-direction average
spectrum of 600 profiles extracted along the $x$-direction. The
spectrum clearly exhibits two regimes. At small wavenumbers ({\it
i.e.} large length scales), a power law behavior is observed with a
slope in the log-log plot of -2 which is consistent with a roughness
exponent of $\zeta_1=0.5$. At large wavenumber ({\it i.e.} small
length scales), a second power law behavior is observed with a larger
slope (-3.2) in agreement with a roughness exponent $\zeta_2=1.1$. The
crossover length scale is sharp and defines a characteristic length
scale $\lambda_c\approx 1$mm. The average spectrum of profiles taken
along the perpendicular direction ($y$-direction) provides a very
consistent result and confirms the isotropy of the surface in terms of
scaling invariance. We checked in Ref.~\cite{Renard03} that mechanical
profiles show the same properties specially at large wavenumbers since
they are sampled at a higher resolution and allow an extended
description of the unusual high roughness exponent $\zeta_2=1.1$.  We
also checked that another analysis technique, namely the Average
Wavelet Coefficient technique \cite{Simonsen98}, was providing very
consistent results. We also show in Ref.~\cite{Renard03} that this
behavior can be observed in several other stylolite samples. The
difference between samples comes from the crossover length scale that
can be larger or smaller, changing drastically the aspect of the
surface. However, the two power law behaviors are conserved with the
same values of the roughness exponents.

The second part of the letter is devoted to a modeling of the
stylolite roughening in order to understand the origin of the
self-affine behaviors and the parameter sensitivity of the
characteristic length $\lambda_c$. As usually done, we reduce the
problem to a 2D situation $z(x)$, {\it i.e.} invariance along the
$y$-direction.

We propose a phenomenological approach that follows the work reported
in {\it Kassner et al.} \cite{Kassner01} who studied the dynamics of a
strained solid in contact with its melt and more specifically the
Asaro-Tiller-Grinfeld instability. In the case of a solid/fluid
interface, the chemical potential between the solid and the fluid at
the boundary can be written as \cite{Asaro72,Gal98}:
\begin{equation}
\label{eq_mu}
\Delta\mu =  \Omega (u_e + \gamma\kappa) 
\end{equation}
where $u_e$ is the elastic energy per unit volume in the solid,
$\gamma$ is the surface energy, $\kappa$ the curvature and $\Omega$ a
molecular volume. We have assumed that gravity effects are negligible. If a
bulk diffusion holds in the fluid, the evolution of the interface is
directly related to the chemical potential: $v_n=m\Delta\mu$ where
$v_n$ is the normal velocity and $m$ is the mobility.

Following the work of {\it Grinfeld and Hazzledine} \cite{Grinfeld96},
we propose to extend the approach to the case of a solid/solid
interface. Moreover, we shall also assume the presence of a fluid in
the pores. Porosity of the material is supposed to be sufficiently
high for having both a bulk diffusion in the fluid and capillary
effects. Doing so, Eq.(\ref{eq_mu}) holds. Since a complete match
between both solid surfaces is assumed, the normal velocity of the
interface is now driven by the relative dissolution between the two
solids.

The difficult task is to expand the local elastic energy along the
interface that takes into account the surface corrugations. Using the
representation theorem for the displacement field and the Hook's law,
it is possible to include the influence of non local elastic
interactions in the limit of small perturbations of the interface
through the elastic Green function as \cite{Bilby68,Gao89}:
\begin{equation}
\sigma(x)=\sigma_0\left(1+\frac{1}{2\pi}PV\int_{-\infty}^\infty{\frac{z(x')
-z(x)}{(x'-x)^2}dx'}\right)
\end{equation}
where $\sigma(x)$ is the local stress along the interface between the
solid, $\sigma_0$ is the average uniaxial external stress or an
effective stress that includes the fluid pressure, and $PV$
is the principal value. According to Ref.~\cite{Kassner01}, the
elastic energy can be approximated for plain strain by:
\begin{equation}
u_e \approx \frac{(1-\nu^2)}{E}\sigma^2(x)
\end{equation}
where $E$ is an effective Young's modulus and $\nu$ an effective
Poisson coefficient.

The last aspect of the modeling concerns a description of the noise
that exists in the chemical potential because of material
fluctuations. We assume that this noise fluctuates on scales
significantly smaller than the scales where elasticity and capillary
effects are considered as for a Langevin approach
\cite{Barabasi95}. However, unlike classical use of a Langevin
equation, we assume here that the noise $\eta(x,z(x))$ is quenched to
represent spatial material fluctuations, and known to have a strong
influence on the scaling properties. For instance, for the Edwards
and Wilkinson (EW) problem \cite{Edwards82,Roux94}, the roughness
exponent is 0.5 in the case of an annealed noise and 1.2 for a
quenched noise.

In the present framework, the interface evolution is described by the
following equation:
\begin{eqnarray}
\label{eq_lang}
\frac{1}{m}\frac{dz}{dt} &=& \Omega
\frac{(1-\nu^2)}{E}\sigma_0^2\left(1+\frac{1}{\pi}PV\int_{-\infty}^\infty{\frac{z(x')
-z(x)}{(x'-x)^2}dx'}\right)\nonumber \\&&+\ 
\Omega\gamma\frac{d^2z}{dx^2}\ +\ \eta(x,z(x))
\end{eqnarray}
which includes three driving forces: the long range elastic
interactions, the local capillary effect which has a stabilizing influence,
and the noise fluctuations. A crossover length scale is obtained
by the balance of the average elastic and capillary terms
in Eq.(\ref{eq_lang})
\cite{Kassner01}:
\begin{equation}
\lambda^* =\frac{\gamma E}{\sigma_0^2(1-\nu^2)}
\end{equation}
Typical values for limestones are $E=8\cdot 10^{10}$ N/m$^2$, $\nu=0.25$,
and $\gamma=20$ J/m$^2$. If we consider a stress at 2 km depth
($\sigma_0=40$ MPa), we obtain $\lambda^*\approx 1$ mm.  We propose
that the $\lambda^*$ length scale is the crossover length scale
$\lambda_c$ observed in Fig.~\ref{fig_spum_data}.

Actually, the mechanical regime ($\lambda\gg\lambda_c$) and the
capillary regime ($\lambda\ll\lambda_c$) have both been extensively
explored but independently. Indeed, the capillary regime is nothing
else than the EW problem in a quenched noise. {\it Roux and Hansen}
\cite{Roux94} have shown in this case that the interface is
self-affine with an exponent $\zeta_2\approx 1.2$. In the mechanical
regime, Eq.(\ref{eq_lang}) is reduced to the quasistatic propagation
of an elastic line in a disordered material. This has been largely
studied \cite{Schmittbuhl95c,Ramanathan98,Rosso02} and the roughness
exponent has been shown to be $\zeta_1\approx 0.4$.

We performed a numerical simulation of Eq.(\ref{eq_lang}) using an
even driven algorithm \cite{Schmittbuhl95c,Schmittbuhl99} where there
is no time evolution. The periodic interface is discretized in N=2048
cells. At start, the interface is flat. At each step, we are searching
for the cell that exhibits the maximum speed $dz/dt$ according to
Eq.(\ref{eq_lang}). This cell is then advanced by a random amount $dz$
uniformly sampled in $[0,1]$. The local fluctuation of the chemical
potential $\eta(x,z(x))$ is updated from a prescribed distribution
chosen here to be uniform between $[0,1]$. Doing so, we are always
dealing with the most unstable cell and let the interface grow there.

\begin{figure}[h]
\begin{center}
\includegraphics[width=8cm,angle=0]{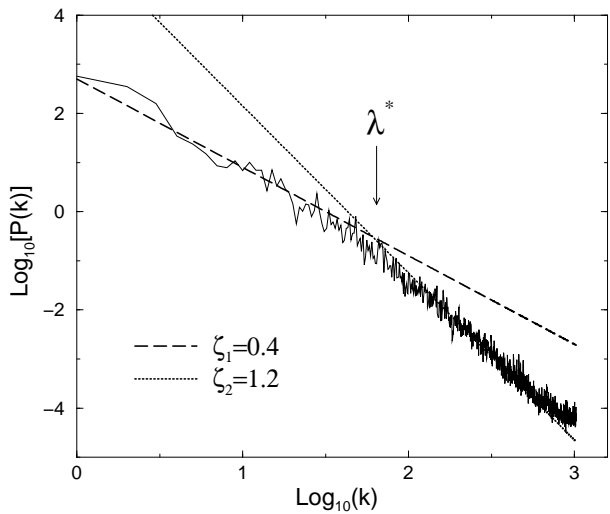}
\caption{\label{fig_spum_simu} Averaged power spectrum of the
topographic profiles obtained from the 2D stylolite modeling.}
\end{center}
\end{figure}

After a transient regime, we observed a stationary width of the rough
interface that develops. The power spectrum of the interface position
$P(k)$, averaged over 500 samples, is shown in
Fig.~\ref{fig_spum_simu}. It reveals a behavior very similar to the
stylolite measurements: A characteristic length scale defined as the
cross-over between two power law regimes. The crossover length scale
$\lambda^\star$ is controlled by the balance between the magnitude of
the mechanical and capillary effects and is compared to the
charateristic length scale $\lambda_c$. As expected, the mechanical
regime with an exponent $\zeta_1\approx 0.4$, is dominating at small
wavenumbers ({\it i.e.}  large length scales). On the contrary, at
large wave numbers, the capillary regime dominates with a roughness
exponent close to $\zeta_2=1.2$. Roughness exponents obtained from the
modeling are slightly different from the measurements. This might be
explained by the dimension difference: experimental surfaces are full
3D interfaces, on the contrary the model is 2D. A complete 3D
numerical modeling is on-going to account for this situation. However,
3D computations are much heavier because of the long delay to reach
the stationary regime of the interface growth.

In conclusion, we presented a quantitative description of stylolite
interfaces that is consistent with a model of interface growth. The
experimental measurement is a high resolution profiling of a 3D
stylolite topography. We show that the surface is self-affine but with
two regimes. At small scales, the roughness exponent is unexpectedly
high, $\zeta_2=1.1$, and consistent with a capillary dominated
regime. At large scales, the stylolites morphology is controlled by
mechanical effects, {\it i.e.} long range elastic stress
redistribution. In this case the roughening is important with a low
roughness exponent $\zeta_1=0.5$. The modeling is based on the
description of a stress-induced instability previously reported as the
Asaro-Tiller-Grinfeld instability. Such framework provides a
prediction of a characteristic length $\lambda_c$ which is the
crossover between the two scaling regimes.  It is important for
geological implications to note that the characteristic length
$\lambda_c$ is very sensitive to the average stress
$\sigma_0$. Indeed, a measurement of $\lambda_c$ from today roughness
profiling could provide an estimate of the stress magnitude during the
stylolite growth, that is, in the past. Thus stylolites could be
considered as stress fossils.

We acknowledge D. Rothman, J. Rice, A. Lobkovsky, B. Evans,
Y. Bernab{\'e}, B. Goff{\'e}, P. Meakin, and E. Merino for very
fruitful discussions. This work was partly supported by the ACI
``Jeunes Chercheurs'' of the French Ministry of Education (JS), and
the ATI of the CNRS (FR).

\bibliographystyle{prsty}

\end{document}